\documentclass[a4paper,11pt,oneside]{article}
\usepackage{graphicx}				
\usepackage{listings}
\usepackage[colorlinks=true,linkcolor=green,anchorcolor=green,citecolor=green,filecolor=green,menucolor=green,urlcolor=blue,breaklinks=true,pdfhighlight=/P,pdfmenubar=true,pdftoolbar=true,pdfpagelabels=true,pdfstartpage=1,pdfstartview=FitV,pdftitle={Computation Speed of the F.A.S.T. Model},pdfsubject={Computation Speed of the F.A.S.T. Model},pdfauthor={Tobias Kretz},pdfcreator={Tobias Kretz},pdfproducer={Tobias Kretz},pdfkeywords={Pedestrian Traffic, Crowd Dynamics, Simulation, Parallel Computing, F.A.S.T. Model}]{hyperref}	
\usepackage[numbers,sort&compress]{natbib}
\usepackage{hypernat}				
\usepackage[american]{babel}	
\begin{document}
\newlength{\figurewidth}\setlength{\figurewidth}{0.618\textwidth}
%
\title{Computation Speed of the F.A.S.T. Model}
\author{Tobias Kretz \\ \\
PTV Planung Transport Verkehr AG\\
Stumpfstra{\ss}e 1\\
D-76131 Karlsruhe\\ \\
\tt{Tobias.Kretz@ptv.de}}

\maketitle

\begin{abstract}
The F.A.S.T. model for microscopic simulation of pedestrians was formulated with the idea of parallelizability and small computation times in general in mind, but so far it was never demonstrated, if it can in fact be implemented efficiently for execution on a multi-core or multi-CPU system. In this contribution results are given on computation times for the F.A.S.T. model on an eight-core PC.

\end{abstract}

\section{Introduction}
Pedestrians and vehicles alike are extended objects in space. For simulation models \cite{Wiedemann1974,Nagel1996,Chowdhury2000,Helbing2001b,Schadschneider2009,Schadschneider2009b} this means that one has to guarantee a mutual exclusion volume, i.e. that they do not overlapp. This exclusion volume includes more than the mere body or vehicle, but as well a headway whose size increases monotonically with speed \cite{Seyfried2005,Chattaraj2009}. This can be achieved in at least two ways: either all agents compute their next movement step in parallel and potentially emerging conflicts about exclusion volumes are solved afterwards \cite{Kirchner2004}, or the movement is done sequentially: a strategy with which it's easy to prevent conflicts generally. However, the kind of update procedure has a strong influence on the dynamics of the system \cite{Schadschneider1998,Rajewsky1998,Kirchner2004}. Obviously parallel update fits better for parallel computing attempts and as a lucky coincidence parallel update has proven to usually yield more realistic results when physical systems are simulated than sequential update \cite{Rogsch2009,Rogsch2009b}.

The F.A.S.T. model \cite{Kretz2006k,Kretz2006d,Kretz2006f,Kretz2007a,Kretz2008f,Kretz2009} tries to make use of both strategies, as the planning process for the next movement step is done in parallel and the actual movement sequentially to avoid a computationally costly conflict resolution lateron. Additionally computationally costly calculations (like exponential or trigonometric functions) are only made use of in the planning process, while actual movement only consists of very simple commands and calculations.

So, each time step consists of a computationally rather expensive planning phase, where data common to all agents is only read (easy to parallelize) and a computationally cheaper actual motion phase, difficult to parallelize, as it writes to data structures common for all agents. In this contribution results of measurements of the computation time of a first parallelized version of the algorithm are given. 

\section{Technical Details}
The parameters of the F.A.S.T. model were chosen to be $k_S=1.2$ and $k_{other}=0$. In a simulation all agents had the same maximum speed. All calculations were done for all maximum speeds $v_m=1$ to $v_m=5$, but as with $v_m=4$ the fundamental diagram of Weidmann \cite{Weidmann1993,Buchmueller2007} is reproduced quite well \cite{Kretz2007a}, the focus in the following results section is on $v_m=4$.

The computation times given in the next section refer to the simulation of 396 simulation time steps equivalating to 396 simulated seconds.
 
The simulations were carried out on a PC with two Xeon E5320 quadcore processors and 20 GB RAM. The parallelization was done using OpenMP \cite{OpenMP}, and the source code was compiled using the Visual C++ 8 (Visual Studio 2005) compiler. 

Parallel computing was made use of only in the process of choosing a desired cell for each agent:

\lstset{language=C}
\begin{lstlisting}
blocksize=max(min(number_of_agents/cores,32767),1);
#pragma omp parallel num_threads(cores)
  {
    #pragma omp for schedule (dynamic, blocksize)
    for (int i = 0; i < number_of_agents; i++) 
      choose_desired_cell (i);
  }
\end{lstlisting}

\section{Results and Conclusions}
\begin{figure}[htbp]
  \center
	\includegraphics[width=\figurewidth]{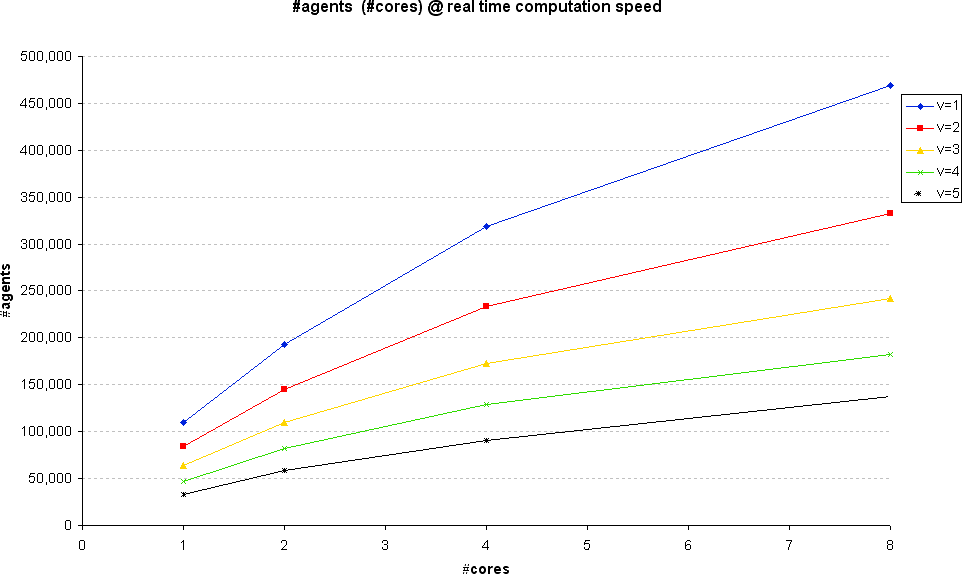}
	\caption{Number of agents that can be simulated in real time in dependence of number of cores.}
	\label{fig:agents_cores}
\end{figure}

\begin{figure}[htbp]
  \center
	\includegraphics[width=\figurewidth]{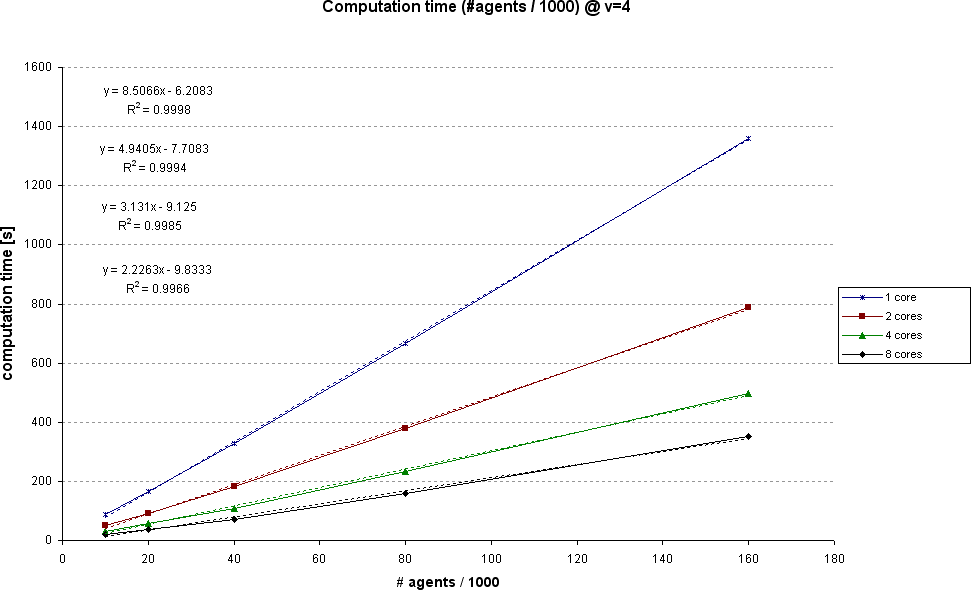}
	\caption{Computation time in dependence of number of agents at fixed maximum speed $v_m=4$.}
	\label{fig:ct_agents}
\end{figure}

\begin{figure}[htbp]
  \center
	\includegraphics[width=\figurewidth]{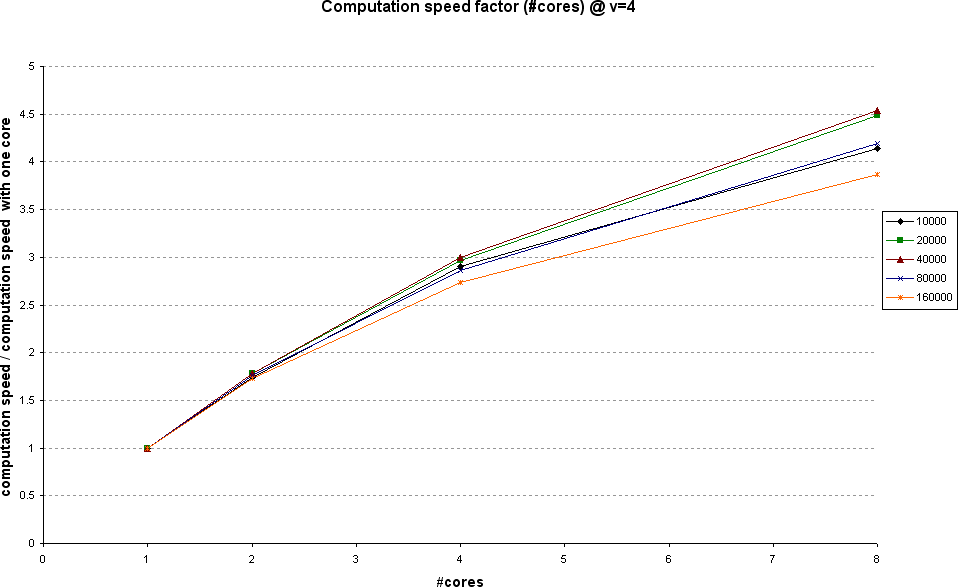}
	\caption{Computation speed factor in dependence of number of cores at fixed maximum speed $v_m=4$.}
	\label{fig:csf_cores_v4}
\end{figure}

\begin{figure}[htbp]
  \center
	\includegraphics[width=\figurewidth]{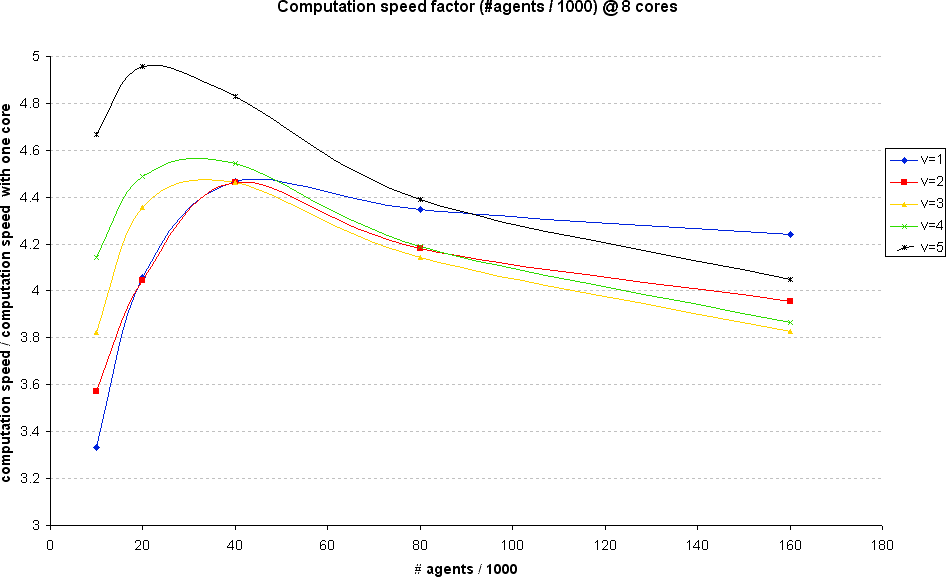}
	\caption{Computation speed factor in dependence of number of agents when eight cores are used (interpolation with splines).}
	\label{fig:csf_agents_8cores}
\end{figure}

\begin{figure}[htbp]
  \center
	\includegraphics[width=\figurewidth]{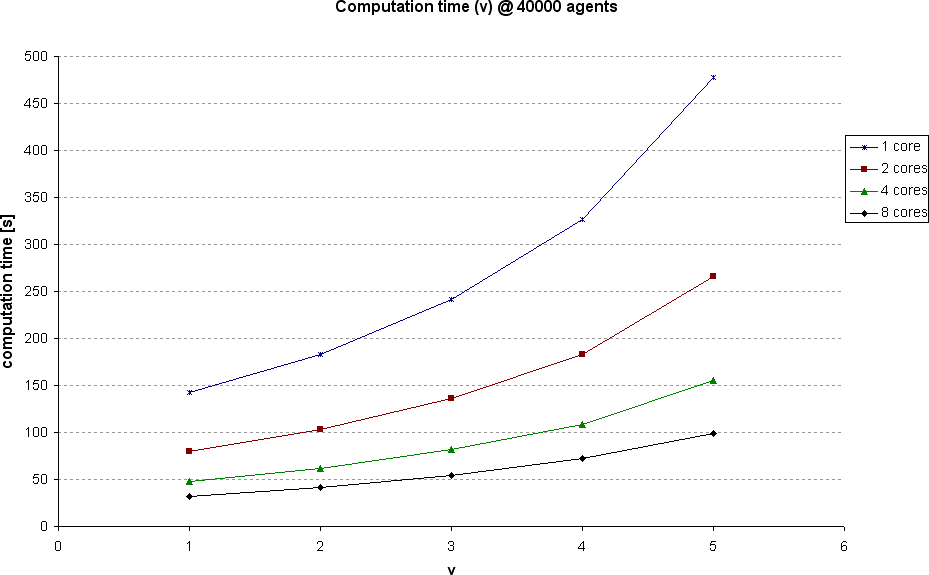}
	\caption{Computation time in dependence of maximum speed for 40,000 agents.}
	\label{fig:ct_v_agents}
\end{figure}

\begin{figure}[htbp]
  \center
	\includegraphics[width=\figurewidth]{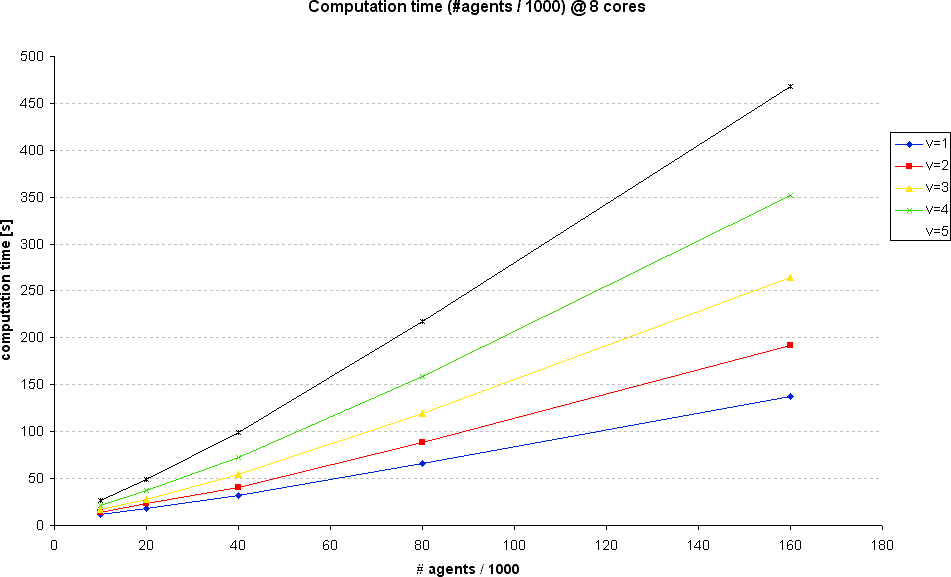}
	\caption{Computation time in dependence of number of agents for 8 cores.}
	\label{fig:ct_agents_cores}
\end{figure}

\begin{figure}[htbp]
  \center
	\includegraphics[width=\figurewidth]{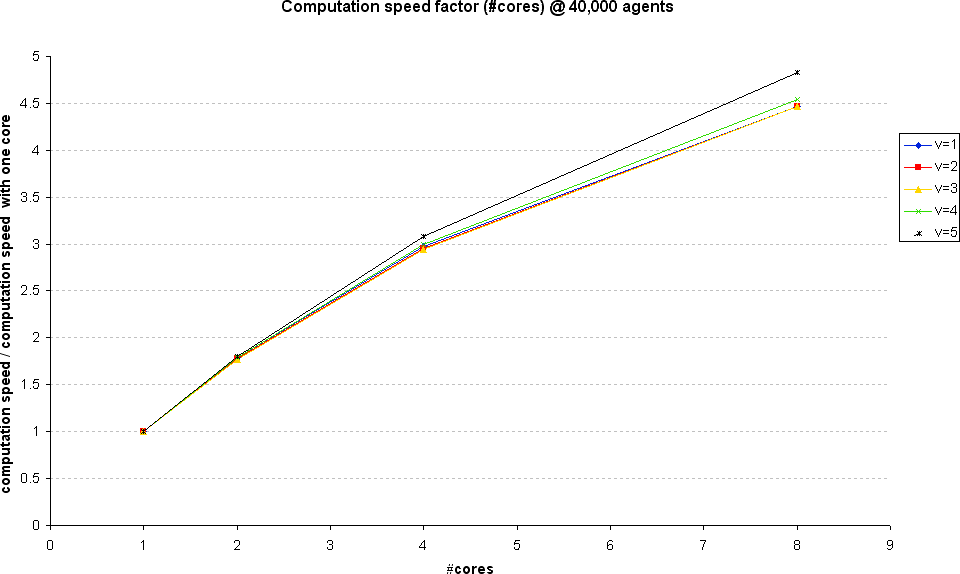}
	\caption{Computation speed factor in dependence of number of cores for 40,000 agents.}
	\label{fig:csf_cores_agents}
\end{figure}

One of the main results from figures \ref{fig:agents_cores} to \ref{fig:csf_cores_agents} is that with a parameter configuration that reproduces Weidmann's fundamental diagram fairly well, and using all eight cores, a real time computation speed could be achieved when about 182,000 agents were in the simulation simultaneously. However, simulating more complex situations like counterflow or non-trivial route choice would require additional calculations and reduce simulation speed.

The computation speed factors with eight cores compared to using only one core of the same PC were found to be in the range 4.3 to 5.0. This confirms that the initial intention is met to have a model suited for parallel computation. 

The wide range of factors in figure \ref{fig:csf_agents_8cores} may be a hint that there might be more efficient partitions of the parallel calculation parts.

Apart from the model efficiently making use of a high number of cores, Intel has released a CPU (Xeon 5482) with a -- depending on the kind of computation -- 20\% to 50\% higher performance. If it is possible to make use of this performance increase, a real-time simulation could be achieved with well beyond 200,000 agents; a stadium size (40,000 agents) evacuation simulation in 5\% to 15\% of real time, as evacuation always implies that the average number of active agents during the course of the simulation is roughly half the initial number.

%
\nocite{_PED2005,_ACRI2006,_ACRI2008,_Chung2009,_PED2008,_Bazzan2009,_Enzy2009}
\bibliographystyle{utphys_quotecomma}
\bibliography{026_TGF09}
%
\end{document}